\documentstyle[12pt]{article}
\def\A{{\rm A}}
\def\N{{\rm N}}
\def\p{{\rm p}}
\def\n{{\rm n}}
\def\NB{{\overline{\rm N}}}
\def\pb{\bar{\rm p}}
\def\nb{\bar{\rm n}}
\def\NNB{\N\hskip -0.5pt\NB}
\def\NBA{\NB\hskip -0.5pt\A}
\def\IM{\mathop{\Im m}\nolimits}
\def\RE{\mathop{\Re e}\nolimits}
\def\etal{{\sl et al.\/}}
\def\apny{{\sl Ann.~Phys.~(N.Y.)\ }}
\def\prl{{\sl Phys.~Rev.~Lett.\ }}
\def\pl{{\sl Phys.~Lett.\ }}
\def\pr{{\sl Phys.~Rev.\ }}
\def\prep{{\sl Phys.~Rep.\ }}

\def\nuph{{\sl Nucl.~Phys.\ }}

\def\v#1{{\bf #1\ }}

 1
 1
 1

\font\tenbf=cmbx10
\font\tenrm=cmr10
\font\tenit=cmti10

\begin{document}
\begin{titlepage}
\begin{flushright}
ISN-95-152\\
nucl-th 9501xxxxxx
\end{flushright}
\vfill\vfill\vfill
\centerline{\tenbf SOME ASPECTS OF ANTIPROTON--NUCLEUS
PHYSICS\footnote{
Invited Talk at LEAP94, Low-Energy Antiproton Physics,
Bled, Slovenia, September 1994}}
\baselineskip=22pt
\baselineskip=16pt
\vspace{0.8cm}
\vfill
\centerline{\tenrm Jean-Marc RICHARD}
\baselineskip=13pt
\centerline{\tenit Institut des Sciences Nucl\'eaires}
\baselineskip=12pt
\centerline{\tenit Universit\'e Joseph Fourier--CNRS--IN2P3}
\centerline{\tenit 53, avenue des Martyrs}
\centerline{\tenit F--38026 GRENOBLE Cedex, France}
\vspace{0.9cm}
\vfill
\begin{abstract}
A brief review of antiproton--nucleus physics is presented.
Some topics are related to early LEAR experiments, and others to more
recent measurements or proposals. These include: exotic molecules,
elastic and inelastic scattering,
 deep annihilation, strangeness production,
neutron--antineutron oscillations, halo nuclei,
antiproton production in nuclear reactions etc.
\end{abstract}
\vfill\vfill\vfil\end{titlepage}
\vspace{0.8cm}
\rm
\baselineskip=14pt
\section{Introduction}
There are many motivations for studying the antinucleon--nucleus
($\NBA$)
interaction. This is first  a  tool
for improving our empirical knowledge of the
``elementary'' nucleon--antinucleon ($\NNB$) process, since 2-body
scattering
cannot yet be studied for all spin and isospin configurations that
are needed for a full reconstruction of the 2-body amplitude.
\par
Conversely, once the behaviour of an antinucleon in ordinary matter
is known,
$\NBA$ experiments can probe some aspects of the nucleus: equation of
state
in a regime of unusual excitation, neutron skin, etc.
\par
Hopefully, $\NBA$ physics is not restricted to a straightforward
folding of the elementary $\NNB$ scattering and annihilation with
wave functions of nuclei.
New phenomena are looked for, such as annihilation on several
nucleons,
increase of  strangeness production, formation of hot bubbles inside
a nucleus,
etc.
\par
Remarkably, the physics topics  related to $\NBA$ experiments
involve different scales, ranging form several tens of fermis, the
typical
Bohr radius for antiprotons, to a fraction of fermi, the distance at
which
quark--antiquark annihilation takes place.
\par
By no means, this review is intended to cover all aspects of $\NBA$
physics. We refer the reader to other contributions, in particular
on induced fission\cite{Egidy_BLED} and Pontecorvo reactions
\cite{Golubeva_BLED}.
\par
\section{Antiprotonic molecules}
There is a renewed interest in antiprotonic atoms.
As shown by Yamazaki at this Conference\cite{Yamazaki_BLED}, the
Coulomb
forces generate metastable states of
systems with a nucleus, an antiproton, and some electrons.
There is already an abundant literature on the
subject\cite{Longlived}. Our
attention was  previously restricted to Rydberg-like atomic states of
an
antiproton around a nucleus, the electrons  either being expelled or
orbiting
far away.
\par
This is a good opportunity to recall that many atomic or molecular
states can be made by combining antiprotons with ordinary nuclei. For
instance, the configurations
\begin{equation}
(\pb\p\p),\qquad(\pb{\rm d}{\rm t}),\qquad(\pb\pb\p\p), \quad\ldots
\end{equation}
 have their ground state which is stable against dissociation
in smaller clusters\cite{ArmB,Ore}.
Some of these exotic states seemingly belong
 to the domain of science fiction. One should remember, however, that
positron or muon chemistry, once at a rather primitive stage, is
nowadays
dealing with rather complex systems.
\par
\section{Antiprotonic atoms and elastic scattering}
The data on elastic scattering and the energy shifts of antiprotonic
atoms
have been successfully analyzed in terms of optical $\NBA$
potentials.
Simple empirical potentials, typically of Wood--Saxon type, have been
tuned to reproduce the data. A moderately attractive real part
is favoured, to supplement the absorptive component.
The optical potential has also been derived by
KMT\cite{KMT,Jensen_BLED}
type of folding of the elementary $\NNB$ amplitude. The resulting
real
potential is usually repulsive inside the nucleus, but it becomes
attractive
near the surface, in the region which is actually probed by
low-energy
scattering or in $\pb$-atoms.
Elaborate medium corrections have been worked out, but they do not
change
the picture too dramatically, as long as the antinucleon interacts at
the
surface.
\par
Microscopic calculations reproduce fairly well the spatial extension
that is
needed to fit the data. While the imaginary potential does not exceed
much the
border of the nuclear density, the real attraction extends a little
outside the
nucleus.
\par
\section{Inelastic scattering}
``Inelastic'' is understood here as in nuclear physics. Namely
one considers a reaction
\begin{equation}
\NB+\A\rightarrow\NB+\A'.
\end{equation}
This includes  charge-exchange processes. The precise identification
of
the new nucleus or nuclear level
A', with known properties, allows one to filter the quantum numbers
which are transferred from the antinucleon to the nucleus, i.e., to
focus on
selected spin and isospin components of the potential.
Extended studies, in particular by Dover and
collaborators\cite{Tignes_Nucleus}, have shown that
this is potentially a very powerful tool to determine the key
characteristics
of the $\NNB$ force, and to check current ideas on the excitation of
nuclear
levels. Preliminary experimental investigations by the
PS$\,184$ collaboration at LEAR\cite{Tignes_Nucleus} were  not
accurate enough
to provide
definite conclusions, and this program has not been resumed with
improved
detectors, unfortunately.
\par
The inclusive $(\pb,\p)$ reaction was also examined, in a search for
possible
$\NBA$ bound states. These states would generalize the $\NNB$
bound states or resonances, which were looked for in  $\NNB$
scattering
and annihilation experiments. The results were
negative\cite{Tignes_Nucleus}.
\par
\section{Integrated aspects of annihilation}
$\NBA$ data confirm that $\NNB$ annihilation is very strong. In
particular,
one needs a very deep absorptive component of the optical potential.
In atoms, the hadronic width $\IM E$ is comparable to the shift $\RE
E$
of the binding energy. We already mentioned the consequence that
low-energy
antiprotons do not penetrate much  into the nuclear medium. Then
comparing
data on neighboring isotopes gives an indication on the
isospin $I=1$ component of the interaction, as the surface of heavier
isotopes is likely to be dominated by neutrons.
\par
An interesting result of $\NBA$ experiments is that annihilation
seems
weaker for $I=1$\cite{Isospin}, i.e.,
\begin{equation}
\sigma_{\rm a}(I=1)<\sigma_{\rm a}(I=0).
\end{equation}
Among the possible theoretical explanations, one may mention
\begin{itemize}
\item[i)] There are less combinatorial possibilities of
quark--antiquark
annihilation in the case of $I=1$.
\item[ii)] For a given intrinsic annihilation strength (for instance
in a
scenario where quark rearrangement would dominate), annihilation
is less effective for $I=1$ than for $I=0$, since the real part of
the
potential is less attractive, at least in models based on meson
exchanges.
According to Shapiro\cite{Sha}, annihilation is monitored by the real
potential, which
focuses the wave function toward the short-distance region.
\end{itemize}
Some results on the isospin dependence are based
on Deuterium or Helium data, by comparing the number
of $(\pb\n)$ annihilation events to that of $(\pb\p)$. In principle,
a
detailed 3-body or 4-body calculations could be carried out, with
phenomenological $\NNB$ potentials.\par
\section{Specific aspects of annihilation}
A recurrent and interesting topics is the so-called $B>0$
 contribution to annihilation, or annihilation on several nucleons.
This means
processes which cannot be reduced to an ordinary $\NNB$ annihilation
followed
by rescattering of the annihilation products. This is discussed in
talks on
Pontecorvo reactions or strangeness production.
\par
Another classic deals with deep annihilation. Higher-energy $\pb$,
those
for instance of the SuperLEAR proposal\cite{SuperLEAR,Torino},  have
a smaller
cross section, and thus penetrate more deeply into the nucleus.
Moreover, the
Lorentz boost
 focuses  the mesons resulting from the primary annihilation in the
forward direction, where they  hardly escape rescattering. Hence a
large energy can be deposited in the nucleus, without the compression
that is experienced in heavy-ion collisions of comparable energy
release. Annihilation of medium-energy antiprotons would never
produce a quark--gluon
plasma, but new types of excited nuclei are sometimes formed, which
probe new
sectors of the equation of state.
\par
At this Conference, however, the fashion is seemingly going backward,
with
more contributions on low-energy than high-energy annihilation. In
particular,
fission induced by $\pb$ has been studied by several groups, with
sophisticated detectors. As explained by von Egidy\cite{Egidy_BLED},
a typical
scenario is the
following: %
\begin{itemize}
\item annihilation at the surface, some fast pions or protons being
emitted.
\item  particle evaporation of the compound nucleus (remember
that in this field, ``particle''means a nucleon or a small
nucleus)
\item binary (sometimes ternary) fission of the compound nucleus
 \item particle evaporation of the fission fragments, etc.
\end{itemize} %
Again, the main interest lies in the comparison with what is observed
in  heavy-ion collisions.
\par
\section{Cold annihilation}
There is a renewed interest in nuclear physics for nuclei with a
neutron
halo. In the simplest case, we have a compact core with $(A-1)$
nucleons,
and a $A^{\rm th}$ nucleon, usually a neutron, with a very small
energy $E$. At
large distances, the wave function behaves like
\begin{equation}
\Psi(\vec{\rm r})\sim\exp(-kr),\qquad
k=\left(-{2mE\over\hbar^2}\right)^{1/2}.
\end{equation}
Estimates give a r.m.s.\ neutron radius exceeding that of the
proton distribution by typically $0.5\,$fm. In the tail, the neutron
density $\varrho_{\n}(r)$ can overcome the proton one,
$\varrho_{\p}(r)$,
by several orders of magnitude. Kaons have been used to probe this
neutron
skin. Antiprotons offer a viable alternative\cite{Golubeva_BLED}, due
to their
large
cross-section. In favorable circumstances, annihilation can take
place
on a neutron at
large distance from the core, so that a cold $(A-1)$ nucleus is
emitted.
There is a proposal for studying such reactions at
LEAR\cite{Polonais}.
\par
Even more interesting are the nuclei with two neutrons in the halo,
say $(\alpha,\n,\n)$, where  $\alpha$ denotes the  core.
Sometimes, one observes the amazing property that $(\alpha,\n,\n)$ is
stable,
while neither $(\alpha,\n)$ nor $(\n,\n)$ is stable. The simplest
case
is $^6$He, i.e., $\alpha=(\p\p\n\n)$. Such nuclei are called
``Borromean'',
after the Borromean rings which are interlaced in such  a subtle
topological
way that if any one of them is removed, the other two become
unlocked.
Borromean binding does not require 3-body forces, and shows up in
simple
Hamiltonian models\cite{FleckJMR}. If for instance one considers
bosons
interacting through the Yukawa interaction
-g$\sum_{i<j}\exp(-\mu r_{ij})/r_{ij}$, the critical coupling $g_3$
for  binding three
particles is  around $20\%$ lower than the critical coupling $g_2$
for 2-body binding. The wave function of such nuclear systems is very
extended,
so if a low-energy antiproton comes in, one could observe a final
state
\begin{equation}
n+\alpha+(\pb\n\rightarrow\hbox{mesons}).
\end{equation}
\par
\section{Neutron--antineutron oscillations}
Proton decay has been predicted in early unified theories
of electroweak and strong interactions, but has never been
seen in underground experiments. Some alternative theories
predict neutron--antineutron  oscillations
($\n\!\leftrightarrow\!\nb$).
A direct, but difficult measurements of $\n\!\leftrightarrow\!\nb$
makes use of high-intensity neutron beams.
An indirect bound on $\n\!\leftrightarrow\!\nb$ is provided by
proton-decay experiments, since the detectors use nuclei
($^{16}$O, $^{40}$Ca, etc.) rather than Hydrogen. The stability of
these nuclei
implies that the protons do not disintegrate, and also that the
neutrons are
not transformed into antineutrons.
Several groups have calculated that the present limit
$T>10^{31}$ years on the stability of matter implies a limit
$\tau>10^8\,$s on the period of $\n\leftrightarrow\nb$. This
calculation
is rather safe, since it mostly uses the $\NBA$ optical potential
near
the surface of nuclei, i.e., precisely in the region where
it has been measured in LEAR experiments. There is a recent
claim\cite{Nazaruk}
 that the bound $\tau>10^8\,$s should be revised by 31 orders of
magnitude,  but it turns out that this new calculation contains an
error\cite{DGR}.
\par
The behaviour of an antinucleon in the nuclear medium is also
probed in more realistic experiments, namely antiproton production
in nuclear collisions\cite{Mosel}.
\par
\section{Conclusions}
Many results on $\NBA$ have been obtained at LEAR and elsewhere, but
the
field is far from being exhausted.
Several new experiments could be done at moderate cost, if the
antiproton
source remains in operation at CERN.
The considerations on medium-energy physics, as developed in the
KAON or SuperLEAR study groups, remain fully valid.
{}From the contributions and discussions at this conference, it is
also clear
that several astute experiments could be done with very cold
antiprotons on selected targets. Radioactive nuclear beams are
routinely
obtained at ISOLDE facility of CERN, and there are new ideas for
getting more
neutron-rich
ions. This is  a theorist's dream to imagine collisions of
antiprotons
with such rare nuclei.
\par
\vspace{0.6cm}
\noindent{\bf Acknowledgments}
 \par
\vspace{0.4cm}
I would like to thank the organizers for the
very stimulating atmosphere of the antiproton conference,
and A.J.\  Cole for comments on the manuscript.
\par\pagebreak
\vspace{0.6cm}
\noindent{\bf References}
 \par
\vspace{0.4cm}
%


\begin{thebibliography}{10}

\bibitem{Egidy_BLED}
T. von Egidy, Talk at this Conference.

\bibitem{Golubeva_BLED}
I.A. Pshenichnov, A.S. Iljinov, and Y.E.S. Golubeva, Talk at this
Conference.

\bibitem{Yamazaki_BLED}
T. Yamazaki, Talk at this Conference.

\bibitem{Longlived}
T. Yamazaki \etal, {\sl Nature} \v{361}(1993) 238;\\ J. Eades, {\sl
Europhysics
  News} \v{24}(1993) 172;\\ K. Richter \etal, \prl\v{66}(1991) 149;\\
W.A.Beck,
  L. Wilets and M.A. Alberg, \pr\v{A48}(1993) 2779;\\ and references
therein to
  earlier works.

\bibitem{ArmB}
E.A.G. Armour and W. Byers Brown, Accounts of Chemical Research, {\bf
26}
  (1993) 168.

\bibitem{Ore}
E.A.~Hylleraas and A.~Ore, Phys.\ Rev.\ {\bf 71} (1947) 493.

\bibitem{KMT}
A.K. Kerman, H. McManus, and R.M. Thaler, \apny \v{8}(1959) 551.

\bibitem{Jensen_BLED}
A. Jensen, in {\sl Antiproton--Nucleon and Antiproton--Nucleus
Interaction},
  Proc. Erice School, ed. F. Bradamante \etal\ (Plenum, N.Y., 1990).

\bibitem{Tignes_Nucleus}
See, for instance, several contributions, in {\sl Proc. 3rd LEAR
Workshop},
  Tignes, 1985, ed. U.~Gastaldi \etal, (Editions Fronti{\`e}res,
  Gif-sur-Yvette, 1985).

\bibitem{Isospin}
F. Balestra \etal, \nuph \v{A491}(1989) 572.

\bibitem{Sha}
I.S.~Shapiro, \prep\v{C35}(1978) 129.

\bibitem{SuperLEAR}
P. Dalpiaz \etal, in {\sl The Elementary Structure of Matter},
Proc.~Les
  Houches Workshop, 1987, ed.~J.-M.~Richard \etal\ Springer-Verlag
(1988).

\bibitem{Torino}
See, for instance, {\it Proc.\ 1st Workshop on Intense Hadron
Facilities and
  Antiproton Physics}, Torino, october 1989, ed. T.~Bressani,
F.~Iazzi and
  G.~Pauli (Italian Physical Society, Bologna, 1990).

\bibitem{Polonais}
J. Jastrzebski \etal, preprint CERN/SPSLC 94-18 (1994).

\bibitem{FleckJMR}
J.-M. Richard and S. Fleck, \prl \v{73} (1994) 1464.

\bibitem{Nazaruk}
V.I. Nazaruk, \pl \v{B337}(1994) 328.

\bibitem{DGR}
C.B. Dover, A. Gal and J.-M. Richard, \pl (in press).

\bibitem{Mosel}
S. Teis, W. Cassing, T. Maruyama, and U. Mosel, \pr \v{D50} (1994)
388.

\end{thebibliography}

\end{document}